\documentclass[aip,jcp,preprint,nofootinbib,amsmath]{revtex4-1}
\usepackage{epsfig,epsf}
\usepackage{graphicx,tabularx}
\newcolumntype{Y}{>{\centering\arraybackslash}X}
\newcolumntype{P}[1]{>{\centering\arraybackslash}p{#1}}
\usepackage{epstopdf}
\usepackage{bm}
\usepackage{color}
\usepackage[labelformat=empty]{subfig}
\begin{document}

\title{Large Slowdown of Water Dynamics at Stacked Phospholipid
  Membranes for Increasing  
  Hydration Level: All-Atoms Molecular Dynamics}

\author{C. Calero}
\email{ccalero@ffn.ub.edu}
\affiliation{Departament de F\'isica Fonamental,
Universitat de Barcelona, Mart\'i Franqu\'es 1, 08028 Barcelona}

\author{H. E. Stanley}
\affiliation{Center for Polymer Studies and Department of Physics,
Boston University, 590 Commonwealth Avenue, Boston, MA 02215, USA}
\author{G. Franzese}
\affiliation{Departament de F\'isica Fonamental,
Universitat de Barcelona, Mart\'i Franqu\'es 1, 08028 Barcelona}

\date{\today}

\begin{abstract}

  Water hydrating phospholipid membranes determine their
  stability and function, as well as their interaction with other
  molecules. In this article we study, using all-atom molecular
  dynamics simulations, the rotational and translational dynamical
  properties of water molecules confined in stacked phospholipid
  membranes at different levels of hydration, from poorly hydrated
  to a completely hydrated membrane. We find that both the
  translational and the reorientation dynamics of water
  are dramatically slowed down as the hydration 
  is reduced. Analyzing in details the structure and dynamics
  of the hydrogen bond at the interface,
  we show that both those among water molecules and those between
  water and
  lipids slow down by decreasing the hydration, however the latter are
  always slower than the former. By increasing hydration, water 
  saturates all the possible hydrogen bonds with the lipids and, by
  further increase of hydration, the hydrogen bonds among waters
  becomes the majority. 
    However, the dynamics of the water-lipids hydrogen
  bonds becomes up to one order of magnitude slower than that of the
  water-water hydrogen bonds, inducing a large slowing down of the
  dynamics of the entire system even at large hydration level.
\end{abstract}



\maketitle




\section{Introduction}\label{Sect__Intro}

Phospholipid membranes provide the framework to biological membranes,
ubiquitous in nature as limiting structures of cells and organelles
which separate interior contents from external environments. They are
also the main component in the formation of liposomes, which are used
as
drug-delivery systems and in cosmetics \cite{Hamley}.  

Phospholipid membranes consist of two leaflets of amphiphilic lipids
which self-assemble due to the hydrophobic effect
\cite{Nagle_BiophysActa2000}. The study of pure component membranes
can help understand how basic biological membranes function and how
they 
interact with the environment. It is of particular interest to
comprehend the properties of water hydrating phospholipid membranes,
since it determines the stability and function of the membrane and
because water
mediates the interaction between membranes and
solutes such as ions, proteins, DNA and other membranes
\cite{Berkowitz_chemrev2006}.  

A range of experimental techniques have been used to study the
properties of water hydrating membranes and other biomolecules,
including neutron scattering \cite{Fitter_JPhysChemB1999}, NMR
\cite{Mazza_PNAS2011,  Wassall_BiophysJ1996}, infrared spectroscopy
\cite{Righini_PRL2007}, ultrafast vibrational spectroscopy
\cite{Zhao_Fayer_JACS2008}, and terahertz spectroscopy
\cite{Tielrooij_BiophysJ2009}. In Ref.~\cite{Wassall_BiophysJ1996} the
diffusion of water confined in the lamellar phase of egg
phosphatidylcholine was investigated as a function of the hydration of
the membrane. An important reduction (approximately a factor 10) of
the water diffusion coefficient was reported for weakly hydrated
systems, which was attributed to the interaction with the membrane;
accordingly, the authors observe a monotonic increase of the diffusion
coefficient with the membrane's hydration.  Recent experiments
\cite{Zhao_Fayer_JACS2008, Tielrooij_BiophysJ2009} have investigated
the dynamical rotational properties of water molecules at phospholipid
membranes with varying water content using different choline-based
phospholipids and experimental techniques. These studies show a global
slow down in the dynamics of water molecules at the interface with
phospholipid membranes and distinguish different types of water
molecules according to their rotational dynamics---fast, irrotational,
and bulk-like---which are related to the number of hydrogen bonds
formed by water molecules. A valid support to the experiments to
interpret the results and understand the properties of hydrated
phospholipid membranes comes from 
all-atoms  molecular dynamics (MD) simulations~\cite{Rog_ChemPhysLett2002, Bhide_JCP2005,
  Berkowitz_chemrev2006}. In particular, Zhang et
al.~\cite{Zhang_Berkowitz_JPhysChemB2009} and Gruenbaum et
al.~\cite{Gruenbaum_JChemPhys_2011} used MD simulations to investigate
the rotational properties and the infrared spectra of water in
hydrated dilauroylphosphatidylcholine (DLPC) and
dipalmitoylphosphatidylcholine (DPPC) bilayers, respectively.  

Here, we study via all-atom MD simulations the dynamical properties
(both diffusion and rotational dynamics) of hydration water at stacked
dimyristoylphosphatidylcholine (DMPC) phospholipid membranes as a
function of their hydration level $\omega$, defined as the  number of
water molecules per phospholipid.  Among a wide variety of lipids,
DMPC are phospholipids incorporating a choline as a headgroup and a
tailgroup formed by two myristoyl chains. Choline based phospholipids
are ubiquitous in cell membranes and used in drug targeting liposomes
\cite{Hamley}. We also relate the dynamical behavior of water confined
between bilayers with the structure and dynamics of the hydrogen bond
network formed with other water molecules and with selected groups of
the lipid.

\begin{figure}
\begin{center}
\vspace*{0.5cm}
\includegraphics*[angle=0, width=12.0cm]{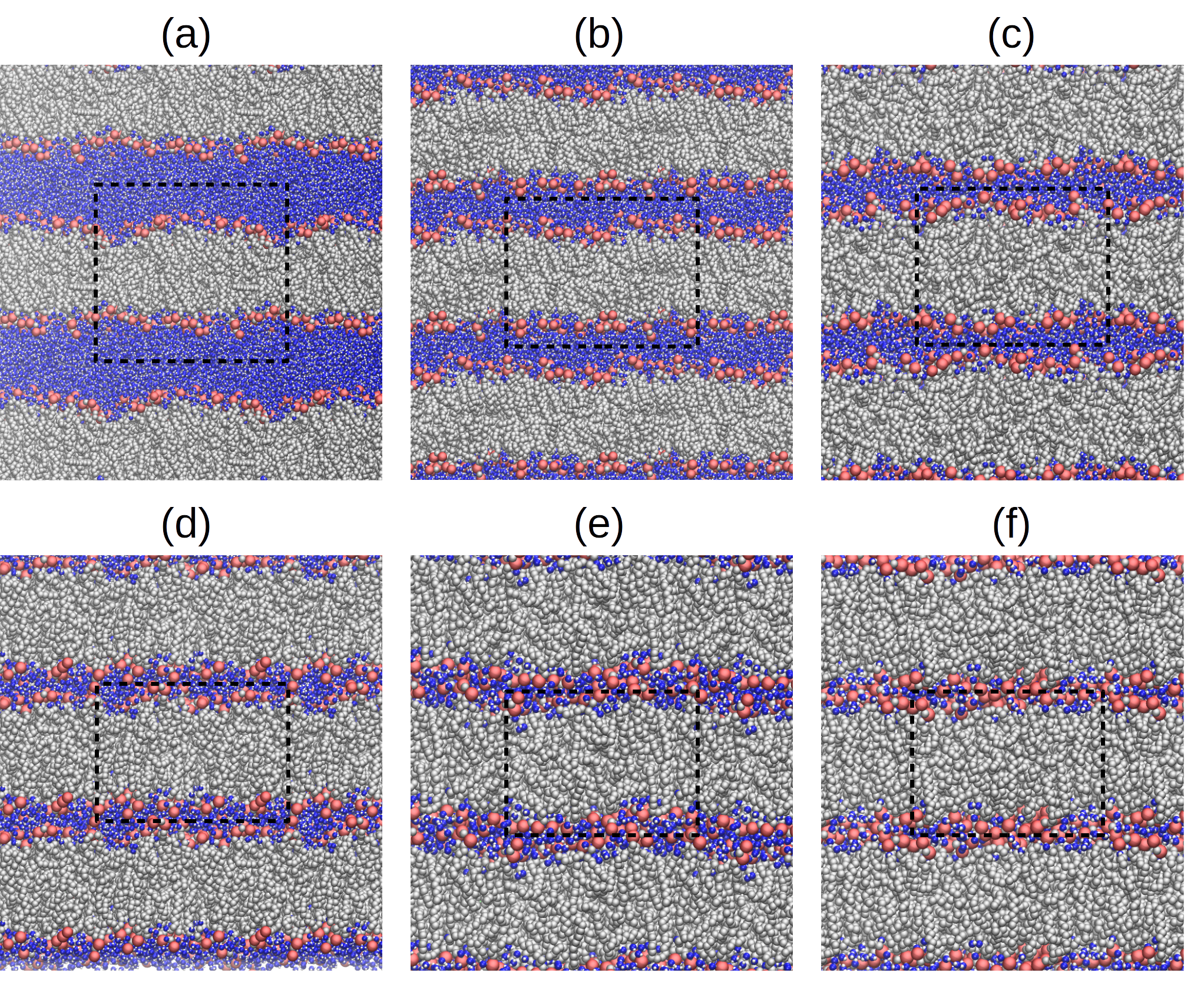}
\caption{Snapshots of the six systems considered in our study, with
  hydration levels (a) 34, (b) 20, (c) 15, (d) 10, (e) 7, and (f)
  4. Gray and red beads represent phospholipid tails and headgroups,
  respectively. Blue and white beads represent oxygen and hydrogen
  atoms of water. The dashed line indicates the size of the simulation
  box. } 
\label{Fig:snapshots}
\end{center}
\end{figure}

\section{Results and Discussion}


We investigate the properties of TIP3P water confined between
phospholipid membranes using MD simulations of hydrated DMPC
phospholipid bilayers (see Methods section for further details).
The use of periodic boundary conditions in our
simulations allows us to describe a system of perfectly stacked
phospholipid bilayers with a homogeneous prescribed hydration level
$\omega$. We consider six different
systems of stacked hydrated phospholipid bilayers, with
hydration levels $\omega =$ 4, 7, 10, 15, 20, and 34 (Fig.~\ref{Fig:snapshots}), which include
the low hydrated systems probed in experiment \cite{Righini_PRL2007,
  Zhao_Fayer_JACS2008, Tielrooij_BiophysJ2009} and a fully hydrated
membrane (with $\omega = 34$) \cite{Nagle_BiophysJ1996}. 

\subsection{Translational dynamics}
\begin{figure}
\begin{center}
\vspace*{0.5cm}
\includegraphics*[angle=0, width=7.0cm]{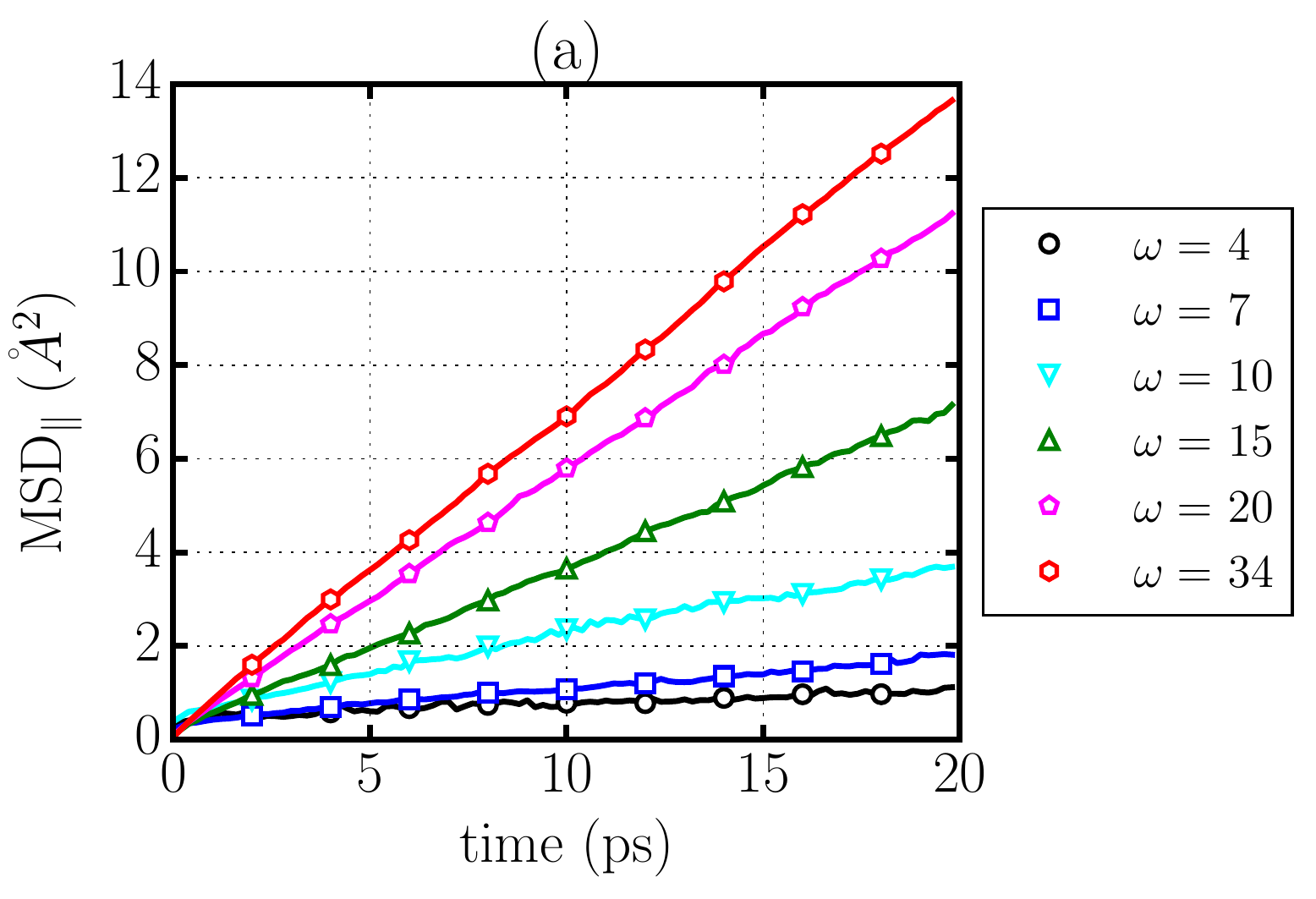}
\includegraphics*[angle=0, width=7.0cm]{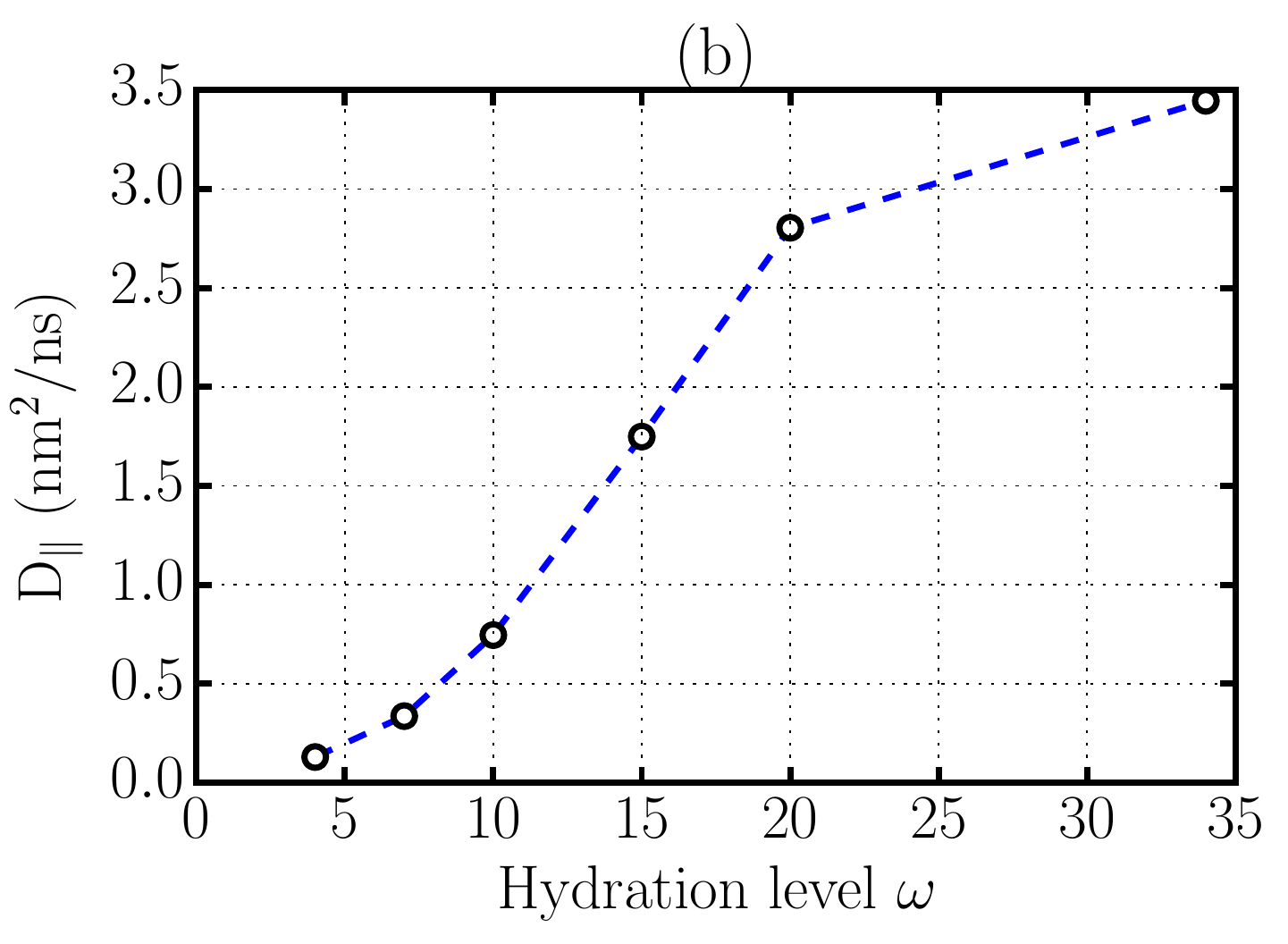}
\caption{Translational dynamics of confined water molecules projected on the
  plane of the membrane for the different stacked phospholipid
  bilayers. (a) Mean-square displacement on the plane of the membrane
  (MSD$_{\parallel}$) as a function of time. (b) Diffusion coefficient
  of water molecules on the plane of the membrane for the different
  hydration levels considered. } 
\label{Fig:diffcoeff}
\end{center}
\end{figure}

We characterize the translational dynamics of water confined in
stacked DMPC bilayers by calculating the mean-square displacement of
the molecule's center of mass projected on the plane of the membrane
(MSD$_{\parallel}$). We simulate 10ns-trajectories saved every 0.1 ps,
discard the initial 2ps until  the diffusive regime is reached, and calculate
the diffusion coefficients for confined water as
\begin{equation}
D_{\parallel} \equiv \lim_{t \to \infty} \frac{\langle |{\bf r}_{\parallel}(t) - {\bf r}_{\parallel}(0)|^2   \rangle}{4t}\,,
\end{equation}
where ${\bf r}_{\parallel}(t)$ is the projection of the center of mass
of a water molecule on the plane of the membrane and the angular
brackets $\langle ... \rangle$ indicate average over all water
molecules and time origins (Fig.~\ref{Fig:diffcoeff}(a)).

We find (Fig.~\ref{Fig:diffcoeff}) that the lower the hydration level
the higher the water slow down, in agreement with previous
experimental and computational studies \cite{Zhao_Fayer_JACS2008,
  Tielrooij_BiophysJ2009, Zhang_Berkowitz_JPhysChemB2009,
  Gruenbaum_JChemPhys_2011}. Indeed, 
the diffusion coefficient increases
monotonically with hydration (Fig.~\ref{Fig:diffcoeff}(b)), from $D_{\parallel} = 0.13$
nm$^2$/ns for the lowest hydrated system ($\omega= 4$) to
3.4 nm$^2$/ns for the completely hydrated membrane
($\omega= 34$), always being much smaller than the bulk value
$D^{bulk} = 5.5$ nm$^2$/ns \cite{Vega_FaradayDisc2009}.
The large drop of the diffusion coefficient
$D_{\parallel}$ for low hydrated membranes with respect to bulk water
and its
dependence with hydration are in qualitative agreement with
experimental results of similar systems \cite{Wassall_BiophysJ1996,
  Rudakova_AppMagRes2004}. Nevertheless, a quantitative comparison with experimental
results might be problematic due to the difficulty in experiment to
ensure the homogeneity of hydration and the perfect alignment of the
membranes in measuring $D_{\parallel}$.  
Moreover, 
this result can be also considered consistent with recent theoretical
advances in our understanding of dimensionality dependence of
diffusion \cite{Seki2016} if we consider
that the membrane is generating a rugged energy landscape for the
confined water. Indeed, the theory of diffusion in a rugged energy
landscape shows that the reduction of the diffusion coefficient from
3d to 2d can be of an order comparable to our result ($\sim$38\% in
our case).

\subsection{Rotational dynamics}

\begin{figure}
\begin{center}
\vspace*{0.5cm}
\includegraphics*[angle=0, width=7.0cm]{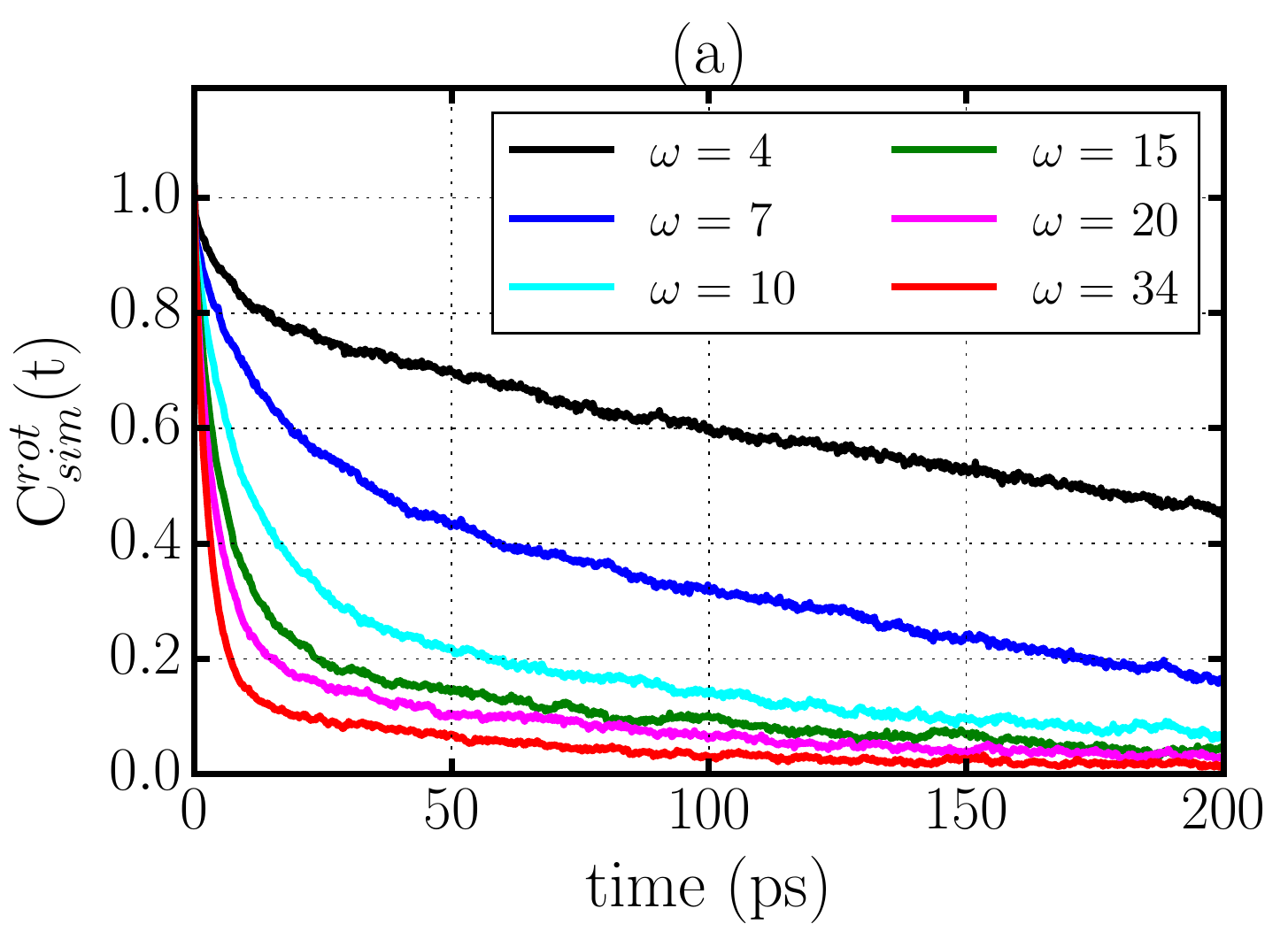}
\includegraphics*[angle=0, width=7.0cm]{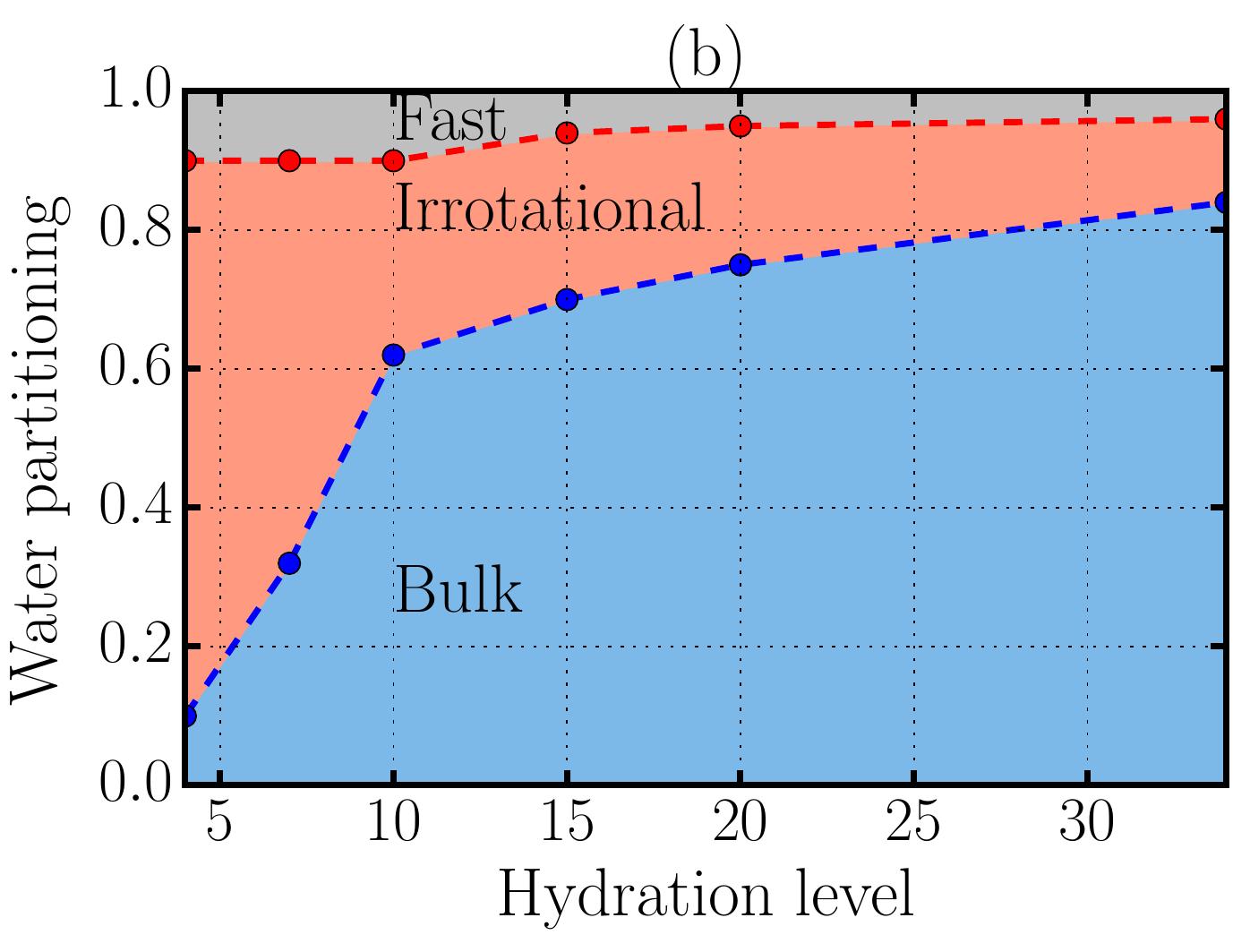}
\caption{(a) Rotational dipolar correlation function of water
  molecules obtained from the simulations of DMPC phospholipid
  bilayers with different levels of hydration $\omega$. (b) Partition
  into (i) bulk-like, (ii) irrotational and (iii) fast water molecules
  as indicated by the fractions  $f_{bulk}$, $f_{irr}$ and $f_{fast}$
  of the three species.} 
\label{Fig:rotcorrfunction}
\end{center}
\end{figure}

Next we study the rotational dynamics of the water molecules
confined in stacked phospholipid bilayers by computing the rotational
dipolar correlation function,  
\begin{equation}
 C_{sim}^{rot}(t) \equiv \langle \hat{\mu}(t)\cdot \hat{\mu}(0) \rangle\,,
\end{equation}
where $\hat{\mu}(t)$ is the direction of the water dipole vector at
time $t$ and $\langle ... \rangle$ denote ensemble average over all
water molecules and time origins (Fig.~\ref{Fig:rotcorrfunction}(a)).
This quantity is related to recent terahertz
dielectric relaxation measurements used to probe the reorientation
dynamics of water \cite{Tielrooij_BiophysJ2009}. To quantify the
relaxation of the correlation functions $C_{sim}^{rot}(t)$ we define
the relaxation time 
\begin{equation}
 \tau_{rot} \equiv \int_0^{\infty} C_{sim}^{rot}(t) dt\,,
\label{tau}
\end{equation}
which is not dependent on any assumptions on the functional form of
the correlation function.
The results for $C_{sim}^{rot}(t)$ 
and the corresponding relaxation times
$\tau_{rot}$ (Table~\ref{table:tauR})
confirm the
monotonic slowing-down  of the rotational dynamics of water for
decreasing membrane hydration as found in
experiments
and in previous computational works \cite{Zhao_Fayer_JACS2008,
  Tielrooij_BiophysJ2009, Zhang_Berkowitz_JPhysChemB2009,
  Gruenbaum_JChemPhys_2011}. 

\begin{table}
\small 
\centering
\begin{tabular}{c|cccccc}
\hline
Hydration ($\omega$) & 4 & 7 & 10 & 15 & 20 & 34 \\
$\tau_{rot}$(ps) & 290 & 99 & 45.6 & 29.0 & 21.7 & 12.4 \\    
\hline
\end{tabular}
\caption{Rotational relaxation time as calculated in Eq.(\ref{tau}) for
  the dipolar correlation function.}
\label{table:tauR}  
\end{table}


Inspection of Fig.~\ref{Fig:rotcorrfunction} suggests the existence of
different time-scales in the decay of the rotational correlation
function $C_{sim}^{rot}(t)$ for all the cases considered: for all
levels of hydration there is an initial rapid decrease of
$C_{sim}^{rot}(t)$ followed by a much slower decay process. To
characterize such time-scales in the relaxation of $C_{sim}^{rot}(t)$
we adopt the conclusions reached in the analysis of experiments by
Tielrooij et al. \cite{Tielrooij_BiophysJ2009} and assume the
existence of three different water species in regards to their
reorientation dynamics: (i) bulk-like water molecules whose
reorientation dynamics resemble that of bulk water, with
characteristic times $\tau_{bulk}$ of a few picoseconds, (ii) fast
water molecules which reorient significantly faster than bulk water
with $\tau_{fast} \approx$ fraction of picosecond, and (iii) irrotational
water molecules which might relax with characteristic times
$\tau_{irr} \gg 10$ps. With this assumption, we can identify the
fractions $f_{fast}$, $f_{bulk}$ and $f_{irr}$ of the three species
as a function of $\omega$
by fitting $C_{sim}^{rot}(t,\omega)$
to a sum of pure exponentially decaying
terms: 
\begin{equation}
 C_{fit}(t,\omega) = f_{fast}(\omega) e^{-t/\tau_{fast}} + f_{bulk}(\omega) e^{-t/\tau_{bulk}} + f_{irr}(\omega) e^{-t/\tau_{irr}}\,,
\end{equation}
with
\begin{equation}
  f_{fast}(\omega) + f_{bulk}(\omega) +
  f_{irr}(\omega) = 1
\end{equation}
for each $\omega$.
Following this procedure we can obtain a measure of
the water partition into the three kinds of water molecules for
different hydration levels of the bilayer. The results of the fits are
shown in Table~\ref{table:partition}, and the corresponding water
partition is represented in Fig.~\ref{Fig:rotcorrfunction} (b). This
diagram qualitatively accounts for the experimental behavior reported
in Ref.~\cite{Tielrooij_BiophysJ2009}. However, such fitting procedure
is not robust--there are five fitting parameters for each
$\omega$--and the parameters $\tau_{fast}$,  $\tau_{bulk}$ and
$\tau_{irr}$ are not showing any regular behavior as function of
$\omega$
(see Table~\ref{table:partition}). This suggests that
the description in terms of such distinctive types of water might not
be complete and that a more thorough analysis is needed.  


\begin{table}
\caption{\label{table:partition} The five parameters of the fits for
  the rotational dipolar correlation function
  $C_{sim}^{rot}(t,\omega)$ resulting from MD
  simulations of stacked DMPC phospholipid membranes with different
  hydration level $\omega$.} 
\small 
\centering
\begin{tabular}{c|ccccc|c}
\hline
$\omega$ & $\tau_{fast}$(ps) & $\tau_{bulk}$(ps) & $\tau_{irr}$(ps) &  $f_{fast}$ &  $f_{bulk}$&  $f_{irr}=1-f_{fast}-f_{bulk}$  \\
\hline
4&0.53 & 5.1 & 321 &0.1 &0.1& 0.8 \\
7&0.48 & 9.8 & 168 &0.1 &0.32& 0.58 \\
10&0.5 & 10.3& 154 & 0.09 &0.62& 0.29\\
15&0.44 & 4.8 & 121 & 0.06& 0.7& 0.24\\
20&0.48& 5.2& 130& 0.05 & 0.75 &0.2\\
34&0.47& 3.2 & 117& 0.04 & 0.85 & 0.11\\
\hline
\end{tabular}
\end{table}

\subsection{Hydrogen bonds structure}

In order to understand the results obtained for the translational and
rotational dynamical properties of water confined at stacked
phospholipid bilayers we analyze the hydrogen bond network
formed by water molecules. In fact, the dynamics and thermodynamics
of liquid water is determined by the breaking and formation of
hydrogen bonds participated by water
\cite{FS2002,  Kumar2006,Laage_Science2006,
  Stokely_PNAS2010,Mazza_PNAS2011,delosSantos2012}. Here, we have 
adopted the widely employed geometric definition of the hydrogen
bond. We consider two molecules to be hydrogen bonded if the distance
between donor and acceptor oxygen atoms satisfies $d_{OO} < 3.5$~\AA~ 
and the angle formed by the OH bond of the donor molecule with the OO
direction is $\theta < 30^o$. We have considered hydrogen bonds formed among water molecules
and also among water and  phosphate or
carboxylate groups of the DMPC phospholipid \cite{Bhide_JCP2005}.  

\begin{figure}
\begin{center}
\vspace*{0.5cm}
\includegraphics*[angle=0, width=8.0cm]{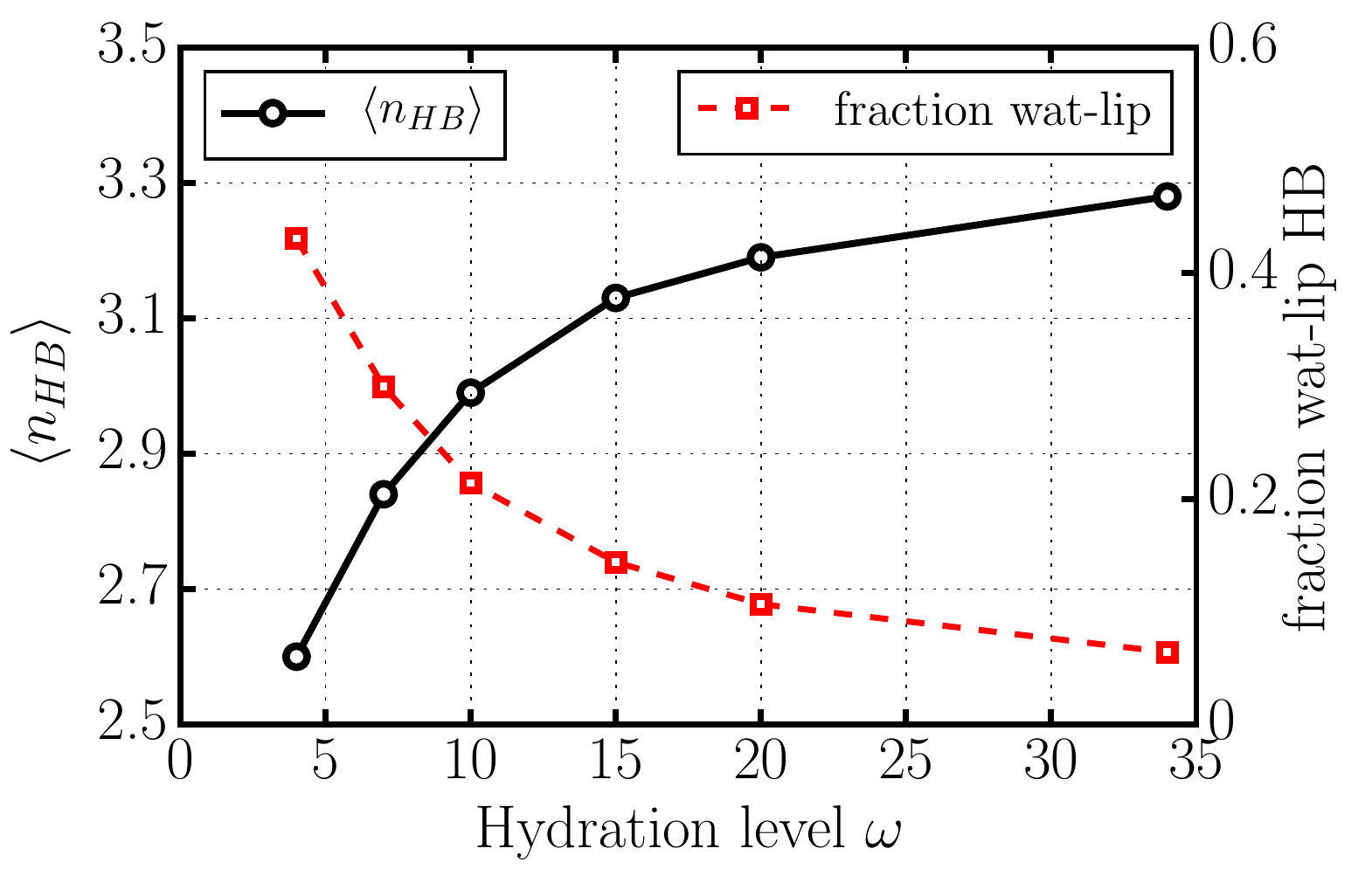}
\caption{Average number $\langle n_{HB}\rangle$ of hydrogen bonds formed
  by water molecules confined in stacked phospholipid membranes as a
  function of their hydration level. We distinguish the average total
  number of hydrogen bonds (black dots),
  and the number of hydrogen bonds of a water molecule
  with lipid groups (red triangles).} 
\label{Fig:avHBhyd}
\end{center}
\end{figure}

Our results show a significant decrease in the average number
$\langle n_{HB}\rangle$ of hydrogen bonds formed by water
as we reduce the membrane hydration $\omega$
(Fig.~\ref{Fig:avHBhyd}).
In addition, the number of hydrogen bonds
formed with selected groups of the lipid increases for decreasing $\omega$
amounting to almost half of
all the hydrogen bonds for the least hydrated case ($\omega = 4$).

\begin{figure}
\begin{center}
\vspace*{0.cm}
\includegraphics*[angle=0, width=7.8cm]{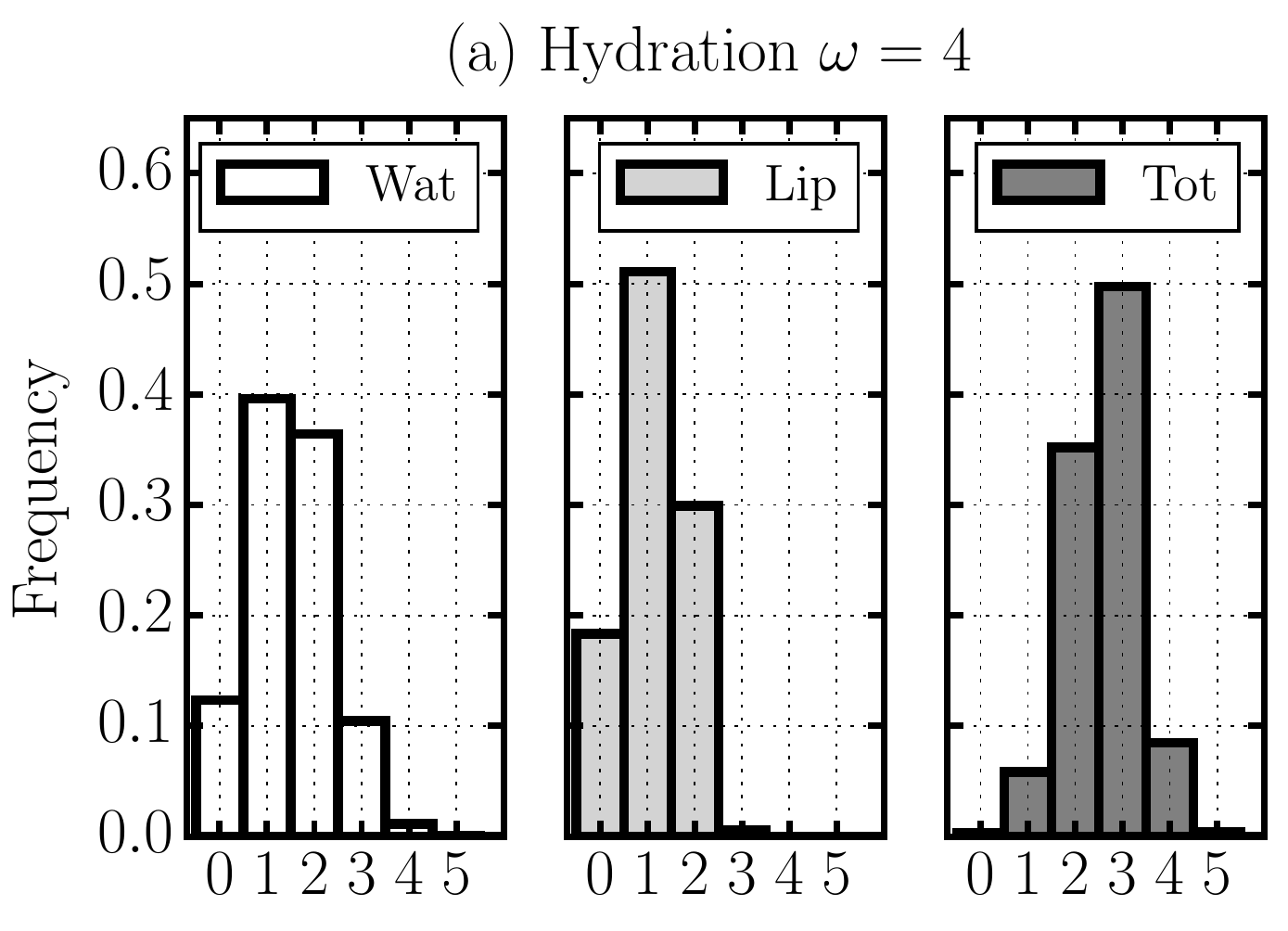}
\includegraphics*[angle=0, width=7.8cm]{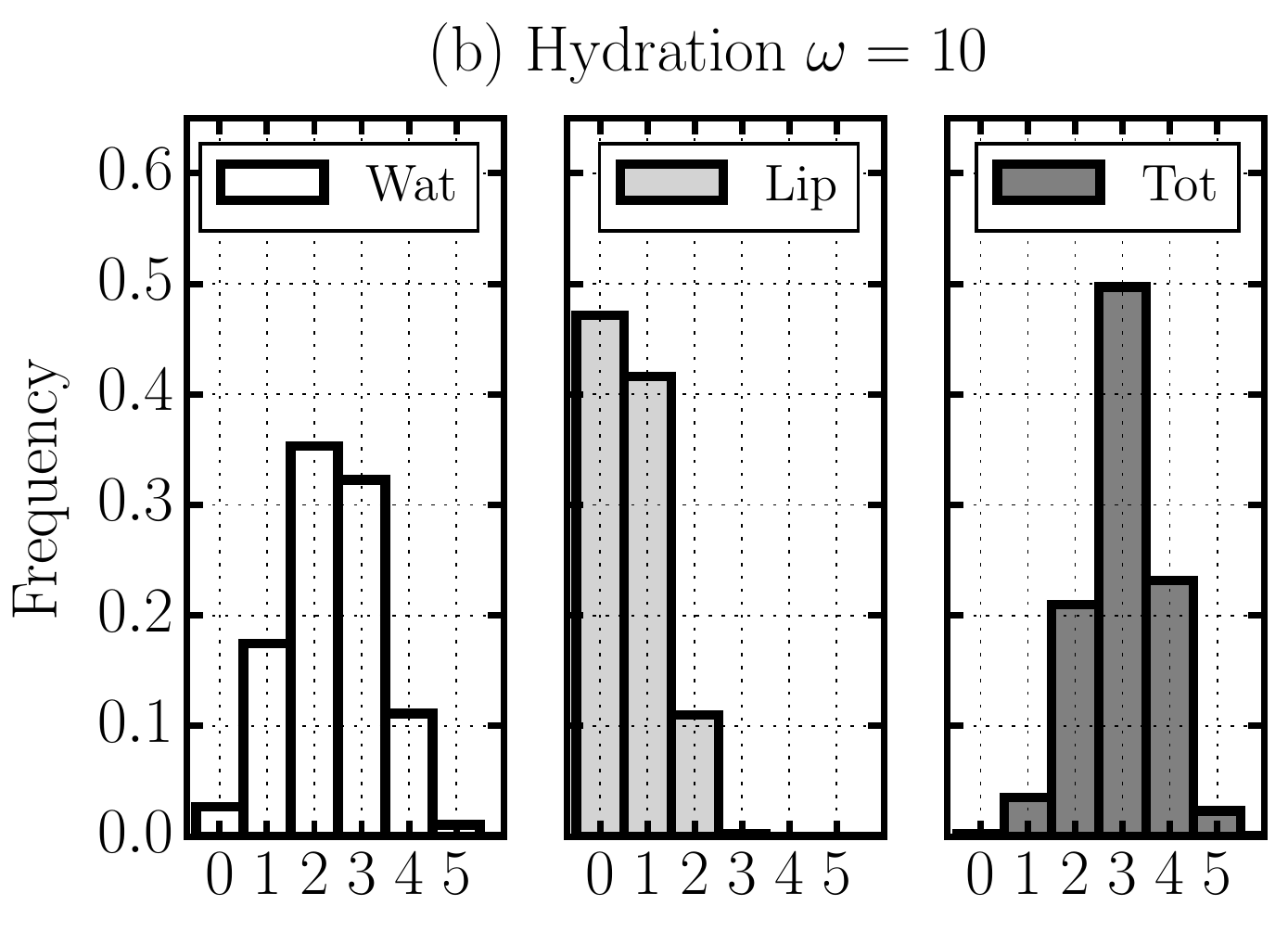}
\includegraphics*[angle=0, width=7.8cm]{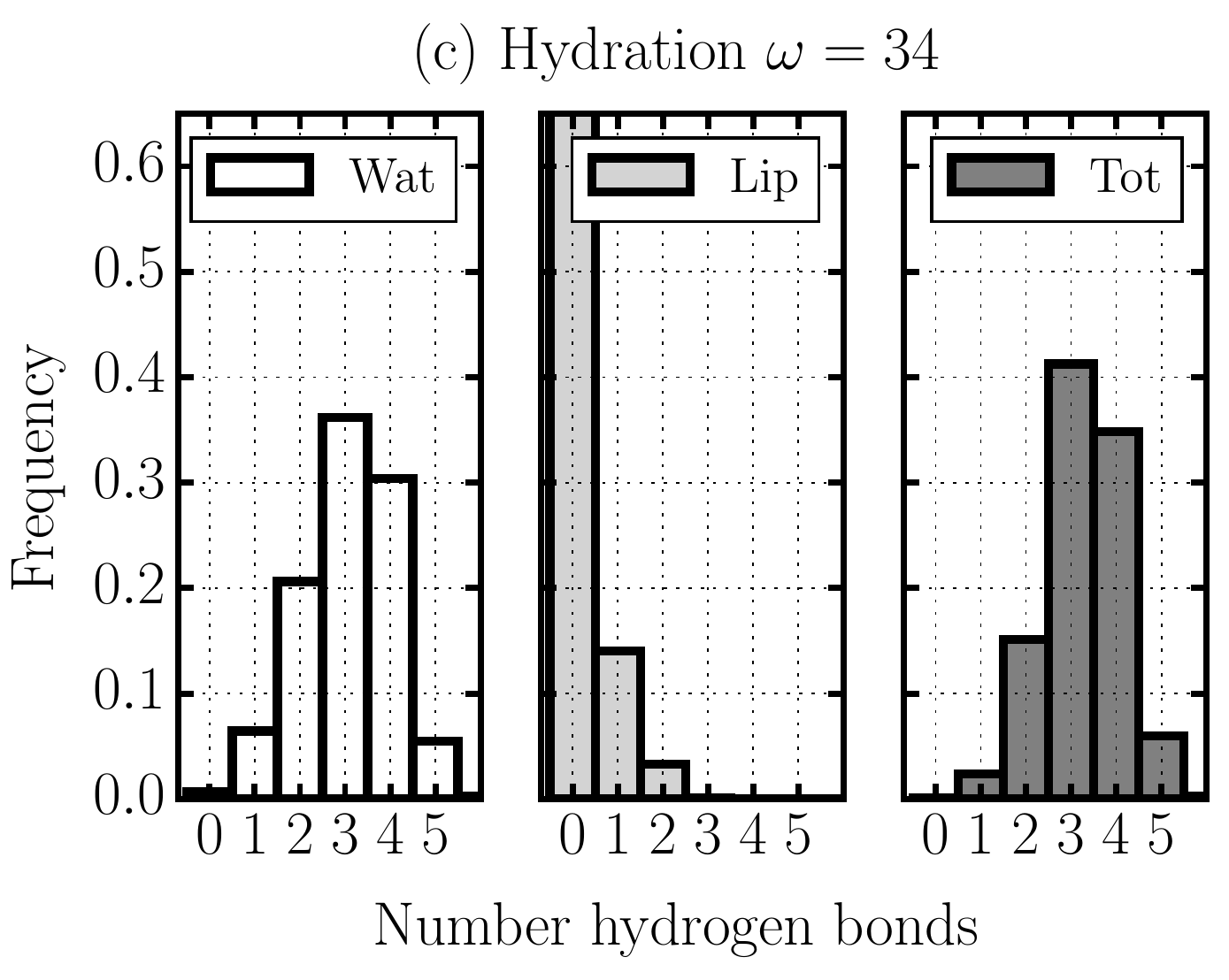}
\caption{Distribution of the number of hydrogen bonds formed by a
  water molecule in stacked membranes with three different levels of
  hydration: (a) $\omega = 4$, (b) $\omega = 10$, and (c) $\omega =
  34$. For each case we show the normalized distribution of the total number of
  hydrogen bonds (right-most panels), 
   of those formed with other water molecules (left-most panels) and of those with
   lipid groups (center panels).} 
\label{Fig:distrHB}
\end{center}
\end{figure}

\begin{figure}
\begin{center}
\vspace*{0.5cm}
\includegraphics*[angle=0, width=12.0cm]{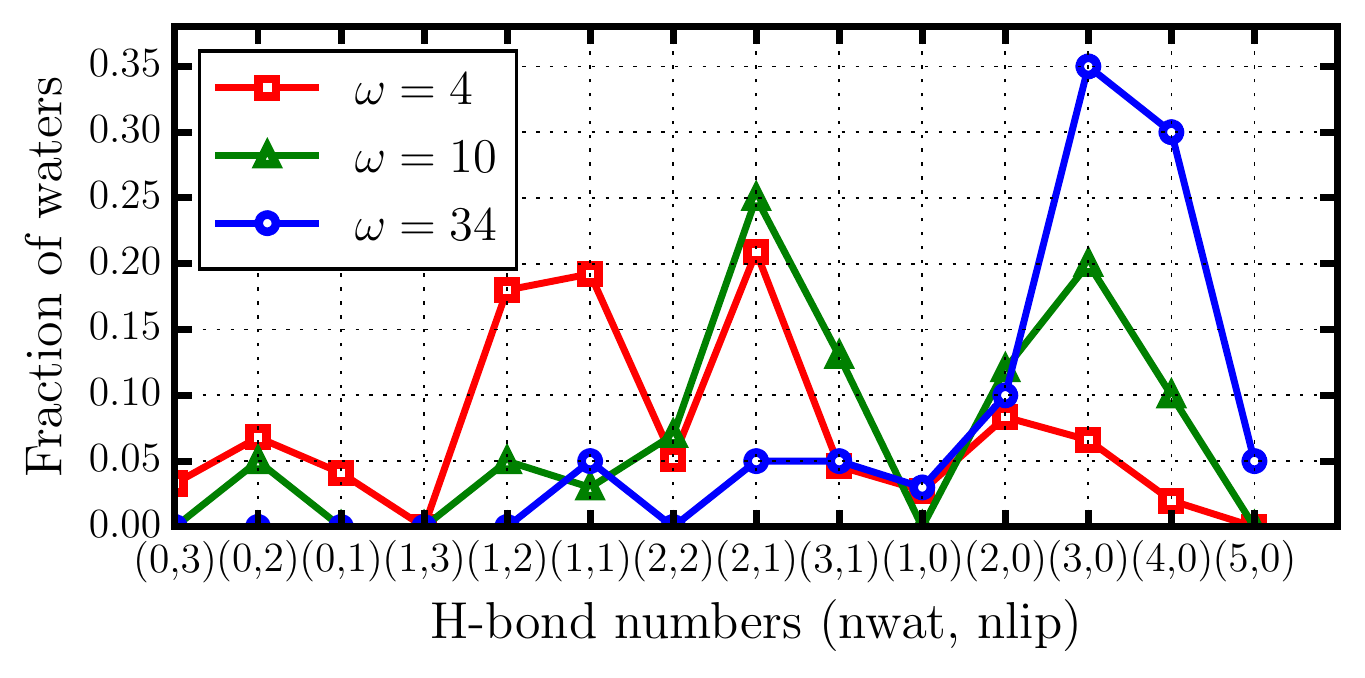}
\caption{Fraction of water molecules forming $n_{wat}$ hydrogen bonds with 
other water molecules and at the same time $n_{lip}$ hydrogen bonds
with lipids, indicated along the $x$-axis as $(n_{wat},n_{lip})$, for hydration  
levels $\omega = 4$, 15, 34. At high hydration (blue circles) the hydrogen bonds are
mainly among water molecules, while at low hydration (red squares) are more likely
those that involve at least one lipid. At medium hydration
(green triangles) we find an intermediate distribution.} 
\label{Fig:HBwatlip}
\end{center}
\end{figure}

To better understand the relation between slowing down of the water dynamics
and structuring of water at the membrane interface, we
analyze how the variation of hydration $\omega$ affects 
the distribution of hydrogen bonds made by water with
water, by water with lipids in the membrane and the distribution of
the total number of existing hydrogen bonds (Fig.~\ref{Fig:distrHB}).
In addition, we calculate the probability distribution that a water
molecule forms $n_{wat}$ hydrogen bonds with other water molecules
{\it and} at the same time 
$n_{lip}$ hydrogen bonds with lipids in the membrane (Fig.~\ref{Fig:HBwatlip}).
We consider in details the results
for the case with lowest hydration $\omega = 4$, with a completely
hydrated membrane $\omega = 34$, and with an intermediate hydration 
$\omega = 10$. We find that the distributions change upon changing the
water content.

For the least hydrated membrane ($\omega = 4$) we find that
a water molecule forms in general two or three hydrogen bonds, of
which with large probability one or two 
with a phospholipid, and one 
or two with another water molecules (Fig.~\ref{Fig:distrHB}(a)).
This implies that the bond configurations with the largest
probabilities---in terms of
$(n_{wat},n_{lip})$---are  
(2,1),
or (1,1), or (1,2), i.e. 
there is a small  probability that a water molecules is not bonded 
at least to one lipid and one water at the same time 
(Fig.~\ref{Fig:HBwatlip}). 

For the membrane with hydration $\omega = 10$ we find that 
a water molecule forms mainly $3\pm 1$ hydrogen bonds (Fig.~\ref{Fig:distrHB}(b)), that are mainly
in the configurations (2,1) and (3,0)
(Fig.~\ref{Fig:HBwatlip}). Therefore, there is a large probability
that a water molecules is bonded to two or three more water molecules,
but there is also a good chance that it will be bonded to a lipid.
%

Finally for the completely
hydrated membrane (with $\omega = 34$) we find that a water molecules
forms more likely three or four hydrogen bonds
(Fig.~\ref{Fig:distrHB}(c)). Of these bonds, very few are with lipids,
while almost all of them are among water molecules (Fig.~\ref{Fig:HBwatlip}).  
%
%

Despite the large differences in the way hydrogen
bonds are formed, the overall hydrogen bond probability distribution
does not change significantly with $\omega$. In all cases, the probability
distribution peaks at three hydrogen bonds and, as the hydration increases,
there is only a shift towards higher numbers of hydrogen bonds of the
secondary peaks. In all cases there is a small probability that a
water molecule forms a single hydrogen bond and this probability 
decreases for increasing hydration level of the
membrane.
By looking at all our results, we suggest that
the population of water molecules with only one hydrogen bond
could represent the fast reorienting water observed in experiment
\cite{Righini_PRL2007, Tielrooij_BiophysJ2009}.

\subsection{Hydrogen bonds dynamics}

\begin{figure}
\begin{center}
\vspace*{0.5cm}
\includegraphics*[angle=0, width=7.0cm]{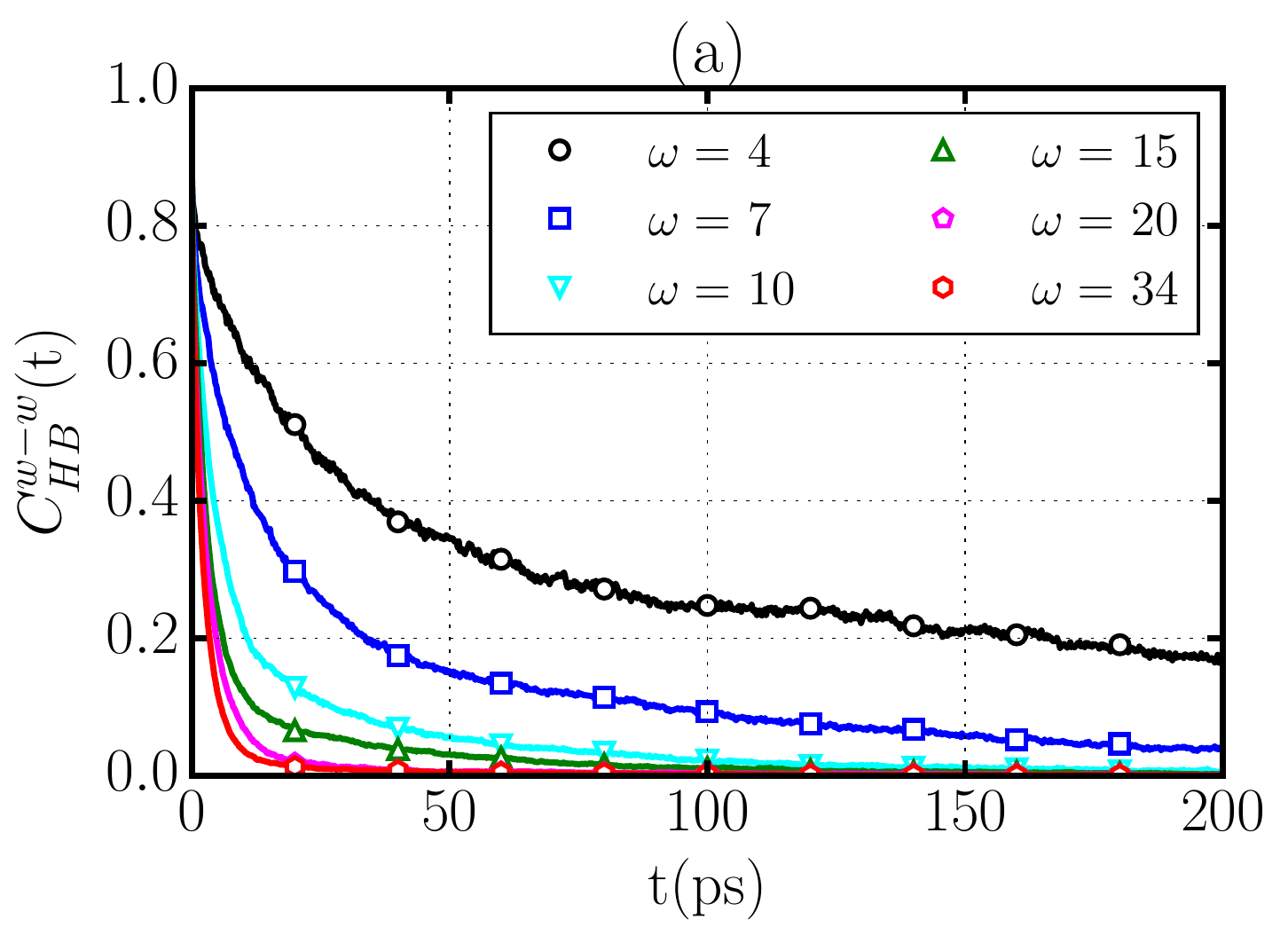}
\includegraphics*[angle=0, width=7.0cm]{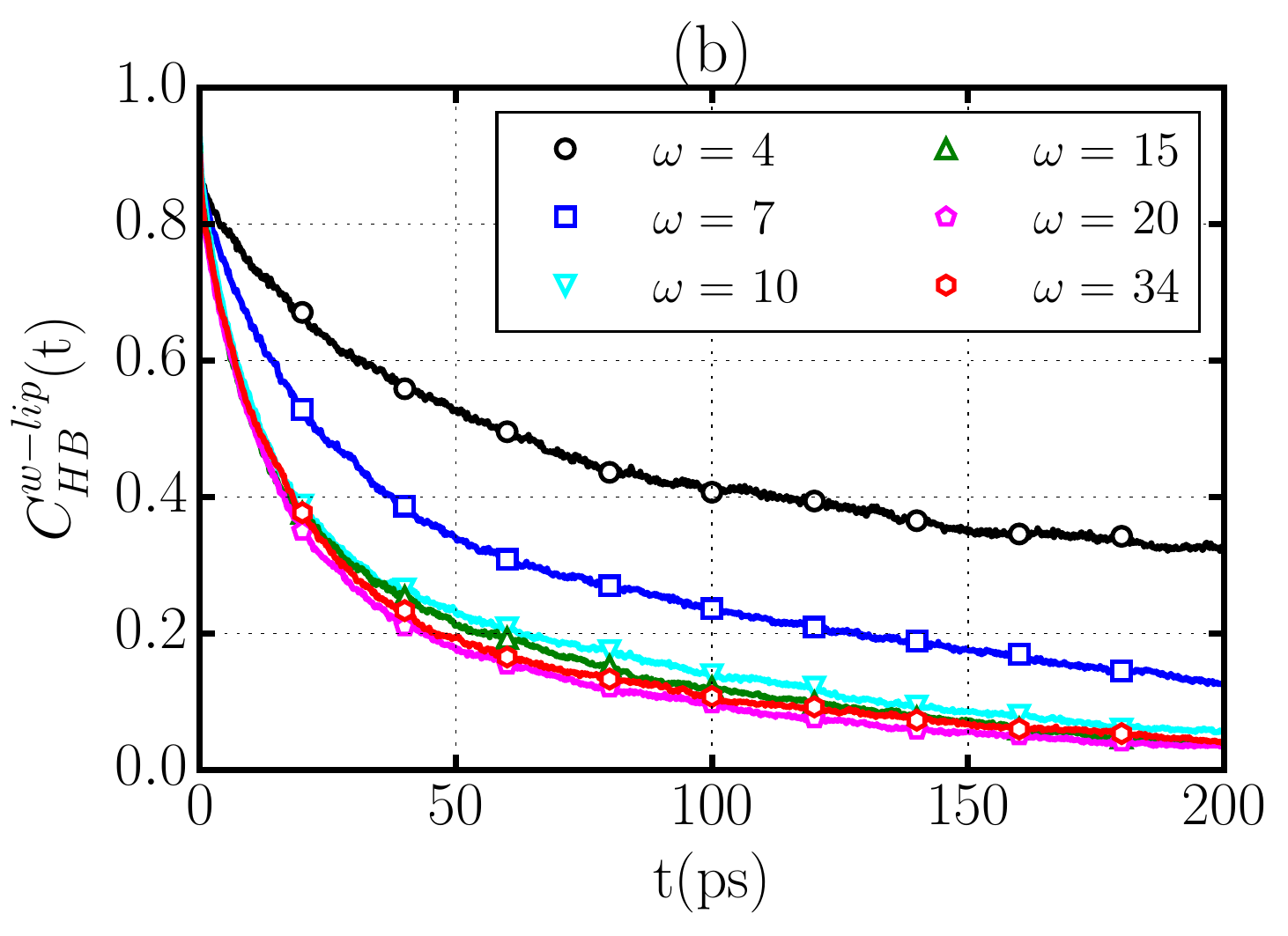}
\caption{Relaxation of the time correlation functions
  $C_{HB}^{w-w}(t)$ (a) and $C_{HB}^{w-l}(t)$ (b) for stacked
  phospholipid membranes with hydration levels $\omega = 4$ (black
  circles), $\omega = 7$ (blue squares), $\omega = 10$ (cyan triangles up),
  $\omega = 15$ (green triangles down), $\omega = 20$ (pink pentagons),
  $\omega = 34$ (red hexagons).} 
\label{Fig:dynamicsHB}
\end{center}
\end{figure}

We have investigated the time evolution of the hydrogen bond network
formed by water molecules to link the results obtained for the
hydrogen bond distribution with the dynamics of confined water at
stacked DMPC phospholipid membranes. To this end, we calculate the
time correlation functions 
\begin{equation}\label{Eq:HBcorrf}
C_{HB}^{w-\alpha}(t) \equiv \frac{ \langle
  n^{w-\alpha}(t)n^{w-\alpha}(0) \rangle}{\langle n^{w-\alpha}(0)
  \rangle}\,, 
\end{equation}
where $n^{w-\alpha}(t)\equiv $ equals one when at time $t$ a given water 
forms a hydrogen bond with another 
water ($\alpha =  w$) or a lipid 
($\alpha = l$), 
and is zero otherwise. The brackets
$\langle ... \rangle$ indicate averaging over all water-water or
water-lipid group pairs and multiple time
origins. $C_{HB}^{w-\alpha}(t)$ provides a measure of the probability
that a hydrogen bond at time $0$ remains formed at a later time $t$
(Fig.~\ref{Fig:dynamicsHB}).

To quantify the relaxation of the correlation functions
$C_{HB}^{w-\alpha}(t)$ we define the relaxation times 
\begin{equation}
 \tau_{HB}^{w-\alpha} \equiv \int_0^{\infty} C_{HB}^{w-\alpha}(t) dt\,,
\end{equation}
which are computed directly from the MD simulations
trajectories. We find that the
relaxation of the water-water hydrogen bond correlation function
$C_{HB}^{w-w}(t)$ slows down dramatically as the hydration level of
the membrane decreases (Fig.~\ref{Fig:dynamicsHB}(a)). In fact, as shown
 in Table~\ref{table:tauHB}, 
the relaxation times of the water-water hydrogen bond network decrease
monotonically from a value of $\tau_{HB}^{w-w} = 110$ ps for the case
with $\omega = 4$ to $\tau_{HB}^{w-w} = 4.0$ ps for the completely
hydrated membrane (with $\omega = 34$). This behavior indicates that
the hydrogen bonds formed between water molecules in proximity to
phospholipids are significantly more stable than those formed between
water molecules in bulk.
This result can be attributed at least to two causes:
\begin{enumerate}
\item  the
formation of hydrogen bonds with water molecules which are in turn
bonded through long-lived hydrogen bonds to the phospholipid
(Fig.~\ref{Fig:dynamicsHB}(b)); 
\item
the lower average density of water molecules at
the membrane interface $\bar{\rho}_{w}$ (see
Table~\ref{table:tauHB}), which hinders the hydrogen bonding switch
between water molecules \cite{Laage_Science2006,
  Zhang_Berkowitz_JPhysChemB2009}.
\end{enumerate}

We find that 
 $C_{HB}^{w-l}(t)$ exhibits a much slower relaxation
than $C_{HB}^{w-w}(t)$ for all cases (Fig.~\ref{Fig:dynamicsHB}(b)).
This fact points to the 
robustness of the water-lipid hydrogen bonds with respect to those
among water molecules.

Furthermore, we find  that the relaxation
of water-lipid hydrogen bonds is unaffected by the level of hydration
of the membrane for cases with $\omega \ge 10$
(Fig.~\ref{Fig:dynamicsHB}(b)), having relaxation times
$\tau_{HB}^{w-l}$ that reach a stable value of 
$\tau_{HB}^{w-l} \approx 40$ ps regardless of the the hydration level
for $\omega \ge 10$ (Table~\ref{table:tauHB}).
This effect can be emphasized
by calculating  the slowing-down factor
$\kappa  \equiv \tau_{HB}^{w-l}/\tau_{HB}^{w-w}$
of the water-lipid   with respect to the
water-water hydrogen bonds. We find that $\kappa$
changes from $\simeq 2$
at low hydration to $\simeq 10$ at fully hydration
(Table~\ref{table:tauHB}).
This result suggests that for $ 7 < \omega < 10$
there is a saturation of the water-membrane interface where water
molecules directly interact with lipid headgroups, in agreement with
X-ray scattering experiments \cite{Kuvcerka_BiophysJ2005}. Upon
increasing the level of hydration, such region of the interface is not
modified and its relative slowing-down with respect to the dynamics of
the water away from the interface become stronger at higher hydration.

Therefore, our results show that
the dynamic slowing-down (both translational and rotational)
of water confined between stacked DMPC phospholipid bilayers
upon decrease of the hydration level is a consequence of a combination
of two factors: (i) the slowing-down of both water-water and
water-lipid hydrogen bonds dynamics, and (ii) a higher proportion
of water-lipid hydrogen bonds at low hydrations.
Moreover, the fact that the effect is strong also at high hydration,
where the proportion of water-water hydrogen bonds is higher, is a
consequence of the fact that water saturates the interface region
forming water-lipid hydrogen bonds that are one order of magnitude
slower than those among water molecules.

\begin{table}
\caption{\label{table:tauHB} Relaxation times of the correlation
  functions $C_{HB}^{w-\alpha}(t)$, $\tau_{HB}^{w-\alpha}$, with
  $\alpha = w, l$, for stacked phospholipid membranes with hydration
  level $\omega$. For each system we  provide the value of the
  average water density, given by $\bar{\rho}_{w} \equiv
  N_{w}/(A_{XY}\times d)$, where $N_{w}$ is the number of water molecules in
  the system, $A_{XY}$ is the area of the membrane, and $d$ is the
  width of the water-membrane interface. We also explicitly indicate
  the slowing-down factor $\kappa$
  of the $w-l$ 
  with respect to the $w-w$ hydrogen bonds.} 
\small 
\centering
\begin{tabular}{c|cccc}
\hline
Hydration ($\omega$) & $\tau_{HB}^{w-w}$(ps) & $\tau_{HB}^{w-l}$(ps) &
$\bar{\rho}_{wat}$ (g/cm$^3$) & $\kappa\equiv \tau_{HB}^{w-l}/\tau_{HB}^{w-w}$\\
\hline
4 & 110 $\pm 10$ & 242 $\pm 10$ & 0.21 & 2.2\\
7 & 31 $\pm 5$ & 80 $\pm 10$ &  0.30 & 2.6\\
10 & 12 $\pm 2$& 44 $\pm 5$ &  0.37 & 3.7\\
15 & 7.1 $\pm 0.5$ & 38 $\pm 5$ & 0.45 & 5.3\\
20 & 5.1 $\pm 0.5$ & 35 $\pm 5$ & 0.52 & 6.7\\
34 & 4.0 $\pm 0.5$ & 38 $\pm 5$ & 0.65 & 9.5 \\    
\hline
\end{tabular}
\end{table}

\section{Conclusions}\label{Sect__Conclusions}

We have investigated the dynamical properties of water confined in
stacked DMPC phospholipid membranes with different hydration levels,
from poorly hydrated systems (with $\omega=4$ water molecules per
phospholipid corresponding to approximately one layer of water between
the two membranes, with membrane-to-membrane distance $h\lesssim 0.3$~nm)
to a completely hydrated membrane (with $\omega = 34$, $10\div 15$
confined water layers, $3\lesssim h/$nm$\lesssim 4.5$). For both the
translational diffusion and the reorientation dynamics of water
we find a dramatic slowing down upon reducing the hydration level.
Indeed, the diffusion coefficient of water on the
plane of the membrane monotonically decreases from 
$D_{\parallel} = 3.4$ nm$^2$/ns for the completely hydrated membrane
to $D_{\parallel} = 0.13$ nm$^2$/ns for the lowest hydrated case.
We find a similar behavior for the water reorientation
dynamics, for which the characteristic relaxation times
increase monotonically from $\tau_{rot} = 12.4$ps at
$\omega = 34$ to  $\tau_{rot} = 290$ps at $\omega = 4$.

To better understand the origin of the slowing down, we analyze how
water organizes near the membrane,
defining the interface as the region where water can form 
hydrogen bonds with the lipids.
We find that the fraction of
water-lipid hydrogen bonds increases for decreasing hydration, while
the water-water hydrogen bonds increase in number for high hydration,
with a relatively small population of water-lipid hydrogen bonds.

The study of the hydrogen-bonds dynamics give us further insight into
the relative proportion of hydrogen bonds between water and lipids and
those among water molecules. Our results show that the population of
the water-lipid hydrogen bonds saturates for increasing hydration and
is characterized by relaxation times that can be one order of
magnitude longer than those for the water-water hydrogen bond. As a
consequence, although at high hydration the water-lipid hydrogen
bonds  are relatively low in number with respect to the total number
of hydrogen bonds, their large increase of characteristic relaxation
time induces a significant slowing-down of the dynamics of the entire
confined water. We emphasize this effect by calculating the slowing-down
factor of the water-lipid hydrogen bonds with respect to water-water
hydrogen bond slowing-down factor.  We observe the consequences of
this slowing-down on the diffusion
constant $D_{\parallel}$  that is strongly reduced with respect to the
bulk case. We partially attribute this reduction to the
heterogeneity of the interface that generates a rugged energy
landscape for the confined water, consistent with recent theoretical
works \cite{Seki2016}. 

In addition, we show that water-water
hydrogen bonds are more robust the lower the hydration of the system;
we attribute such an effect  (i) to the formation of hydrogen bonds with
water molecules which are in turn hydrogen-bonded to the lipid and
(ii) to
the lower water density at the interface in those systems, which
hinders the hydrogen bonding switch between water molecules.
Both effects contribute to an overall slow down of the dynamics of the
confined water upon reduction of  the hydration of stacked DMPC phospholipid. 


We have also analyzed the reorientation dynamics of water molecules in
terms of the existence of three distinctive types of water molecules
(fast, bulk-like and irrotational), as suggested by recent experiments
\cite{Tielrooij_BiophysJ2009}. From a multiexponential fit of the
water reorientation dynamics we obtain the partition of water
molecules into the assumed water types, which qualitatively agrees
with experiment. However, the best fits are obtained for relaxation
times which are different for each of the hydration levels considered,
which suggests that such a partition might be an incomplete
description of the confined water system and that further analysis is
required.

In conclusion, our analysis clearly show that the formation of
long-living, slow-relaxing hydrogen bonds of water with the lipids at the interface
with the membrane are among the main responsible for the large dynamic
slowing-down of water confined between membranes. Our results are
possibly relevant for studying the mechanical properties of biological
membranes.

\section{Methods}\label{Sect__Simulations}

We prepare six different systems of hydrated phospholipid bilayers
with hydration levels (i.e. water molecules per lipid) $\omega =$ 4,
7, 10, 15, 20, and 34 (Fig.~\ref{Fig:snapshots}); we consider from
the weakly hydrated systems used in experiment
\cite{Zhao_Fayer_JACS2008, Tielrooij_BiophysJ2009, Righini_PRL2007} to
a fully hydrated membrane (with hydration level $\omega = 34$)
\cite{Nagle_BiophysJ1996}. The bilayer is composed by 128
dimyristoylphosphatidylcholine (DMPC) lipids distributed in two
leaflets. We apply periodic boundary conditions in all three
dimensions, which allows us to describe a system of stacked bilayers
with perfect periodicity along the direction perpendicular to the
plane of the membrane.  

We perform Molecular Dynamics (MD) simulations using the molecular
dynamics simulation package NAMD 2.9 \cite{Phillips_JCompChem2005} at
a temperature of 303 K and an average pressure of 1 atm.
We set  the
simulation time step to 1 fs.
We describe the structure of phospholipids
and their mutual interactions  by the recently
parameterized force field CHARMM36 \cite{Klauda_JPhysChemB2010,
  Lim_JPhysChemB2012}, which is able to reproduce the area per lipid
in excellent agreement with experimental data. The water model
employed in our simulations, consistent with the parametrization of
CHARMM36, is the modified TIP3P \cite{Jorgensen_JCP1983,
  MacKerell_JPhysChemB1998}.
We cut off the Van der Waals interactions 
at 12 \AA with a smooth switching function starting at 10
\AA.
We compute the long ranged electrostatic forces with the help of
the particle mesh Ewald method \cite{Essmann_JCP1995}, with a grid
space of about 1~\AA.
We update the electrostatic interactions every 2~fs. After energy
minimization, we equilibrate each system for  
10~ns followed by a production run of 50~ns in the NPT ensemble at 1
atm. In the simulations, we control the temperature  by a Langevin
thermostat~\cite{Berendsen_JPhysChem1984} with a damping coefficient of
1~ps$^{-1}$, and we control the pressure by a Nos\'{e}-Hoover
Langevin barostat~\cite{Feller_JChemPhys1995} with a piston oscillation
time of 200~fs and a damping time of 100~fs.


\vspace{6pt}  

\begin{acknowledgments}
We thank Marco Bernabei, Valentino Bianco, Sergey Buldyrev, Jordi Mart\`i and Oriol
Vilanova for useful discussions.
C.C. and G.F. acknowledge the support of Spanish MINECO grant
FIS2012-31025.
C.C. acknowledges the support from the Catalan Governament Beatriu de Pin\'os program
(BP-DGR 2011).
The author thankfully acknowledges the computer resources, technical
expertise and assistance provided by the Red Espa\~nola de
Supercomputaci\'on.

\end{acknowledgments}


\bibliographystyle{unsrt}
\bibliography{DynamicsWater}

\end{document}